\documentclass[article]{aa}

\usepackage{graphicx}
\usepackage{natbib}
\usepackage{scalerel}

\usepackage[table]{xcolor}

\usepackage{txfonts}
\usepackage[pdfencoding=auto,psdextra]{hyperref}
\hypersetup{
    colorlinks=true,
    linkcolor=blue,
    filecolor=magenta,      
    urlcolor=blue,
    citecolor=blue
}
\makeatletter
\renewcommand*\aa@pageof{, page \thepage{} of \pageref*{LastPage}}
\makeatother

\usepackage[utf8]{inputenc}

\usepackage[switch, modulo]{lineno}

\usepackage{euclid}

\begin{document}
%
%

\title{Quantifying the impact of detection bias from blended galaxies on cosmic shear surveys}

   
\newcommand{\orcid}[1]{} 
\author{
Eray Genc$^{1,2}$\thanks{E-mail: egenc@astro.rub.de}, 
Peter Schneider$^{2}$, 
Sandra Unruh$^{2}$, 
Tim Schrabback$^{3}$}

%
%
%
%

\institute{$^{1}$Ruhr University Bochum, Faculty of Physics and Astronomy, Astronomical Institute (AIRUB), German Centre for Cosmological \\Lensing, 44780 Bochum, Germany\\
$^{2}$Argelander-Institut f. Astronomie, University of Bonn, Auf dem Hügel 71, 53121 Bonn, Germany\\
$^{3}$University of Innsbruck, Institute for Astro- and Particle Physics, Technikerstraße 25, 6020 Innsbruck, Austria
}

%
%
\abstract{Increasingly large areas in cosmic shear surveys lead to a
  reduction of statistical errors, necessitating to control systematic
  errors increasingly better. One of these systematic effects was
  initially studied by Hartlap et al. in 2011, namely that image
  overlap with (bright foreground) galaxies may prevent some distant
  (source) galaxies to remain undetected. Since this overlap is more
  likely to occur in regions of high foreground density -- which tend
  to be the regions in which the shear is largest -- this detection
  bias would cause an underestimation of the estimated shear
  correlation function. This detection bias adds to the possible
  systematic of image blending, where nearby pairs or multiplets of
  images render shear estimates more uncertain and thus may cause a
  reduction in their statistical weight. Based on simulations with
  data from the Kilo-Degree Survey, we study the conditions under
  which images are not detected. We find an approximate analytic
  expression for the detection probability in terms of the
  separation and brightness ratio to the neighbouring
  galaxies. Applying this fitting formula to weak lensing ray tracing
  through, and the galaxy distribution in the Millennium Simulation,
  we estimate that the detection bias alone leads to an underestimate of
  $S_8=\sigma_8\sqrt{\Omega_\mathrm{m}/0.3}$ by almost 2\% and can
  therefore not be neglected in current and forthcoming cosmic shear
  surveys.}
%
%
%
%
   \titlerunning{The effects of blended galaxies on cosmic shear surveys }
   \authorrunning{E. Genc, P. Schneider, S. Unruh \& T. Schrabback}
   
   \maketitle
%
%
%
%
   
\section{\label{sc:Intro}Introduction}
Gravitational lensing refers to the distortion of light from distant
galaxies, as it passes through the gravitational potential of
intervening matter along the line of sight. This distortion occurs
because mass curves space-time, causing
light to travel along curved paths. This effect is independent of the
nature of the matter generating the gravitational field, and thus
probes the sum of dark and visible matter. In cases where the
distortions in galaxy shapes are small, a statistical analysis including many
background galaxies is required; this regime is known as weak
gravitational lensing. One of the main observational probes within
this regime is `cosmic shear', which measures coherent distortions (or
`shears') in the observed shapes of distant galaxies, induced by the
large-scale structure of the Universe. By analysing correlations in
the shapes of these background galaxies, one can infer statistical
properties of the matter distribution and put constraints on
cosmological parameters.

Although the large areas covered by recent imaging surveys, such as
the Kilo-Degree Survey \citep[KiDS;][]{de2013kilo}, significantly reduce
statistical uncertainties in gravitational lensing studies, systematic
effects need to be studied in more detail. One such systematic is the
effect of galaxy blending, which generally introduces two key
challenges: first, some galaxies may not be detected at all; second,
the shapes of blended galaxies may be measured inaccurately, leading
to biased shear estimates. While most recent studies focus on the
latter effect
\citep{hoekstra2017study,mandelbaum2018weak,samuroff2018dark, martinet_2019}, the
impact of undetected sources, first explored by
\citet{hartlap2011bias}, has received limited attention
since. \citet{hartlap2011bias} investigated this detection bias by
selectively removing pairs of galaxies based on their angular separation and
comparing the resulting shear correlation functions with and without
such selection. Their findings showed that detection bias becomes
particularly significant on angular scales below a few arcminutes,
introducing errors of several percent. Given the magnitude of this
effect, the detection bias cannot be ignored -- this serves as the
primary motivation for our study. Although mitigation strategies such
as the \texttt{Metadetection} have been proposed \citep{sheldon2020},
challenges remain, especially in the case of blends involving galaxies
at different redshifts, as highlighted by \citet{nourbak}.

Simply removing galaxies from the analysis \citep{hartlap2011bias}
leads to object selection that depends on number density, and thus
also biases the cosmological inference, for example, by altering the
redshift distribution of the analysed galaxies. While
\citet{hartlap2011bias} explored this effect using binary exclusion
criteria based on angular separation, our work expands on this by
modelling the detection probability as a continuous function of
observable galaxy properties -- specifically, the flux ratio and
projected separation to neighbouring sources. This enables a more
nuanced and physically motivated treatment of blending. Based on this
analysis, we aim to construct a detection probability function that
can be used to assign statistical weights to galaxies, rather than
discarding them entirely, thereby mitigating bias without altering the
underlying redshift distribution.

This paper is structured as follows. In Sect.~\ref{sec:theory}, we
describe the theoretical framework underlying our
study. Section~\ref{sec:data} presents the data used in our
analysis. The methodology to determine the detection probability
function, along with the corresponding results, is detailed in
Sect.~\ref{sec:methods}. We discuss our findings and conclude in
Sect.~\ref{sec:conclusion}.

\section{Theoretical framework}
\label{sec:theory}
The theoretical framework of this work is based on weak gravitational
lensing \citep{bartelmann2001weak}. Shape distortions of background
galaxies are quantified using the shear $\gamma$ and the
convergence $\kappa$, which are combined in the
Jacobian matrix of the lens equation,
\begin{equation}
    \mathcal{A} = \begin{pmatrix} 1-\kappa-\gamma_1 & -\gamma_2  \\-\gamma_2 & 1-\kappa+\gamma_1\end{pmatrix},
\label{matrix:jac}
\end{equation}
where the complex shear is introduced with its two components,
$\gamma=\gamma_1+\mathrm{i}\gamma_2=|\gamma|\,\mathrm{e}^{2\mathrm{i}\phi}$,
with $\phi$ being the phase describing the direction of
distortion. Assuming initially circular sources, the elongation of
their images is described by the shear, which causes them to appear
elliptical, while convergence quantifies the change in size. As
neither convergence nor shear is observable on its own, one
rewrites the Jacobian matrix as
\begin{equation}
    \mathcal{A} = (1-\kappa)\begin{pmatrix}1-g_1 & -g_2\\-g_2 & 1+g_1\end{pmatrix},
  \end{equation}
where we introduce the reduced shear $g$,
\begin{equation}
    g(\mathbf{\theta}) = \frac{\gamma(\mathbf{\theta})}{1-\kappa(\mathbf{\theta})}=|g|\,\mathrm{e}^{2\mathrm{i}\phi},
  \end{equation}
which is a measurable quantity used in weak gravitational lensing. The
observed image ellipticity provides an unbiased estimator of the
reduced shear \citep{schrammkayser,seitzschneider}.

Because most of the distortions are weak and thus difficult to
quantify, multiple statistical tools have been developed. A prominent
example are the two-point shear correlation functions (2PSCFs), which
measure how the shears at the position of the source galaxies are
correlated. The shear is decomposed into two components relative to
the direction $\phi$ of the separation vector of each galaxy pair, the
tangential and cross components, defined as
\begin{equation}
    \gamma_\mathrm{t} = -\mathfrak{R}(\gamma \mathrm{e}^{-2\mathrm{i}\phi}),\hspace{0.6cm} \gamma_\times=-\mathfrak{I}(\gamma \mathrm{e}^{-2\mathrm{i}\phi})\,.
  \end{equation}
  Then the 2PSCFs can be defined as
\begin{equation}
    \xi_\mathrm{\pm}(\theta) = \langle \gamma_\mathrm{t}\gamma_\mathrm{t}\rangle(\theta)\pm \langle \gamma_\times\gamma_\times\rangle(\theta), \hspace{0.6cm} \xi_\times(\theta) = \langle \gamma_\mathrm{t}\gamma_\times\rangle(\theta)\,,
\end{equation}
where $\theta$ is the projected separation between two galaxies.

Measurements of correlation functions are carried out as follows. On a
field of galaxies, all galaxy pairs with an angular separation
$\theta$ are considered. Then the average of the product of tangential
ellipticities, $\langle \epsilon_\mathrm{t,i} \epsilon_\mathrm{t,j}
\rangle$, of the pairs is calculated, which is an estimator of $\langle
\gamma_\mathrm{t}\gamma_\mathrm{t} \rangle$ assuming that the
intrinsic ellipticities are uncorrelated. After executing the same
procedure for the cross component, we can calculate the shear
correlation functions above. The advantage of this method is that it
is not affected by bad pixels or masked regions as the mean shear
vanishes.

It is expected that $\xi_\times$ vanishes due to parity invariance in the Universe and is thus useful to check for some systematics, while the other correlation functions $\xi_\pm$ are related to the power spectrum $P_\kappa(\ell)$ as
\begin{equation}
    \xi_+(\theta) = \int_0^\infty
    \frac{\ell\,\mathrm{d}\ell}{2\pi}\,\mathrm{J}_0(\ell\theta)\,P_\kappa(\ell)\,
    , \hspace{0.4cm} \xi_-(\theta) = \int_0^\infty
    \frac{\ell\,\mathrm{d}\ell}{2\pi}\,\mathrm{J}_4(\ell\theta)\,P_\kappa(\ell)\,, 
\label{eq:corrbessels}
\end{equation}
where $J_0$ and $J_4$ are the zeroth- and fourth-order Bessel
functions of the first kind, respectively, while $P_\kappa(\ell)$
denotes the convergence power spectrum. Correlation functions from a
data field are typically estimated by binning, whose estimator is
written as \citep{schneider2002analysis}
\begin{equation}
    \xi_{\pm}(\theta) = \frac{\sum_{ij}\omega_i\omega_j(\epsilon_{{\rm t},i;j}\epsilon_{{\rm t},j;i} \pm \epsilon_{\times,i;j}\epsilon_{\times,j;i})\,\Delta_{ij}(\theta)}{\sum_{ij}\omega_i\omega_j \, \Delta_{ij}(\theta)}\,,
\label{eq:main_corrfunc}
\end{equation}
where $\omega_{i,j}$ denotes the weights assigned to the galaxies and $\Delta_{ij}(\theta) = 1$ if the separation between galaxies $i$ and $j$ falls within the bin centred on $\theta$, and 0 otherwise.
\section{Data sets}
\label{sec:data}
As mentioned in the previous sections, in this work we aim to quantify
the detection bias resulting from the blended galaxies in cosmic shear
surveys. To this end, we simulate galaxies with
\texttt{GALSIM}\footnote{\texttt{GALSIM} is an open-source image
  simulation tool for creating realistic astronomical images. Its key
  advantage lies in offering a wide range of options for
  transformations, rotations, convolutions, and other image-processing
  operations. For more information the reader is referred to
  \citet{rowe2015galsim}.} and insert them into $r$-band detection
images from the fourth data release of the Kilo-Degree Survey
(KiDS). After source extraction with \texttt{SExtractor} \citep[v2.23.2,][]{bertin_1996},
the resulting catalogue is compared with the
list of simulated sources, allowing us to assess which galaxies are
missed under specific conditions. In this context, `detection' refers
to whether a simulated source is identified by
\texttt{SExtractor}, and the detection probability is defined as the
ratio of detected to inserted sources.

In the second part of this study, we use these detection probabilities
as weights in the computation of shear correlation functions. The
shear data are derived from the Millennium Simulation, complemented by
galaxy properties from the catalogue of
\citet{henriques2015galaxy}. This combined analysis enables us to
assess the impact of detection bias on cosmic shear measurements.

\subsection{Kilo-Degree Survey data}
The Kilo-Degree Survey is an optical imaging survey, which covers
$\sim 7\%$ of the extragalactic sky ($\sim 1350$ deg$^2$) performed
with the OmegaCAM wide-field camera\footnote{OmegaCAM is a wide-field
  camera on the VST with a resolution of 268 Megapixels, a
  field-of-view (FOV) of $1^\circ \times 1^\circ$ and a pixel scale of
  $0.\!\!^{\prime\prime}213/$pix. For more information about OmegaCAM
  the reader is referred to \citet{verdoes2013monitoring}.} on the VLT
Survey Telescope (VST) with a diameter of $2.6\,\mathrm{m}$, which is
located at La Silla Paranal Observatory in Chile operated by European
Southern Observatory \citep[ESO;][]{de2013kilo}.  

In this work, we used data from the fourth data release, KiDS-1000,
which consists of 1006 survey tiles covering over $1000\,{\rm deg}^2$. The source detection, photometry, and general information about the data products of KiDS-1000 can be found in \citet{kuijken2019fourth}, while the reduction of KiDS data is explained in \citet{de2015first}.
Specifically, we use the $r$-band detection images, which are 1800\,s
exposures with a PSF FWHM $\leq 0.\!\!^{\prime\prime}8$ and a mean
airmass of 1.3.

Galaxies with reliable redshift and shape estimates are compiled into
the publicly available KiDS-1000 SOM-gold catalogue,
containing 21\,262\,011 sources. This catalogue provides fluxes,
magnitudes, half-light radii, redshifts, and ellipticities. The
photometric processing is described in \citet{kuijken2019fourth}, and
the redshift calibration -- based on self-organizing maps (SOMs) trained
on spectroscopic samples -- is detailed in \citet{hildebrandt2021kids}
and \citet{wright2020photometric}.

Galaxy clusters in KiDS are identified using the Adaptive Matched
Identifier of Clustered Objects (AMICO) algorithm
\citep{bellagamba18}, which models the observed galaxy distribution as
a superposition of cluster signal and background noise. After
identifying the most significant cluster, AMICO subtracts its signal
to search for additional candidates. The detection is performed on
galaxies with $r < 24$, resulting in a catalogue
of 7988 clusters in the redshift range $0.1 < z < 0.8$ with SNR
$> 3.5$. We used the AMICO-DR3 catalogue, which includes clusters from
KiDS Data Release 3 only. Consequently, our analysis of cluster
environments is limited to $\sim450$ deg$^2$. More details about the
AMICO-DR3 catalogue are available in \citet{maturi2019amico}.

\subsection{Gravitational lensing simulations}
As our study investigates the impact of blended galaxies --
particularly the conditions under which they are detected or missed --
cosmological simulations play a central role. They offer a controlled
environment to systematically explore how detection is influenced by
factors such as galaxy density, brightness, and proximity to
neighbouring sources.

To simulate the mass distribution of the Universe, we used the
Millennium Simulation \citep{springel2005simulations}, which
is a DM-only $N$-body simulation with a number of particles
$N=2160^3$, each having a mass of $m_{\rm p}=8.6\times10^8h^{-1}M_\odot$, in a
cube with a comoving side length of $L=500\,h^{-1}\mathrm{Mpc}$. The
growth of structure is tracked with these discrete DM particles from
redshift $z=127$ to $z=0$ in the framework of $\Lambda$CDM cosmology
with $\Omega_\mathrm{m}=0.25$, $\Omega_\mathrm{b}=0.045$,
$\Omega_\Lambda=0.75$, $h=0.73$, $n_\mathrm{s}=1$, and
$\sigma_8=0.9$. The simulation accurately traces the large-scale
structure and the growth of dark matter halos, providing a robust
framework for connecting cosmology to observational lensing data.

To connect the results of $N$-body simulations to the lensing
observables, such as shear and convergence, we use the ray-tracing
results presented in \citet{hilbert2009ray}. They are based on the
construction of a backward light cone along multiple lens planes from
37 snapshots at redshifts of between $z=0$ and $z=3.06$, each having a
FOV of $4^{\circ}\times4^{\circ}$ corresponding to a $4096^2$ pixel
grid.

As galaxies are the primary observables in weak lensing surveys,
connecting dark matter simulations to real data requires a model of
galaxy formation. To this end, we employ the semi-analytic galaxy
catalogue of \citet{henriques2015galaxy}, which builds on the halo
merger trees of the Millennium Simulation. This model incorporates
baryonic processes such as gas cooling, star formation, and feedback,
with parameters tuned to match observational constraints, primarily
from early SDSS data \citep{stoughton2002sloan}. The result is a
realistic galaxy population that can be used to study how detection
biases propagate into weak lensing measurements.

\section{Methods and results}
\label{sec:methods}
To quantify the effects of blended galaxies, the detection probability
of galaxies is investigated as a function of different parameters. For
this, we simulate galaxies with
\texttt{GALSIM}, insert them into the KiDS-1000 $r$-band detection
images, and examine which sources are detected. In this section, we
present two main methods for that purpose with their results. The
first method investigates the detection probability of galaxies in
galaxy cluster regions as a function of the clusters' redshift and
richness in Sect.~\ref{subsec:4.1}, while the second method examines
how the detection probability behaves depending on different
properties of the inserted sources or their neighbours in
Sect.~\ref{subsec:4.2}. The following procedure is applied to both
methods:
\begin{enumerate}
    \item For a field having $N$ galaxies, we insert $N/10$ simulated galaxies, which are simulated with \texttt{GALSIM} as described in the following. 
    \item A light profile with a Sérsic index $n$ randomly drawn from
      the distribution shown in Fig.~\ref{fig:cosmos_hist}, total flux
      $F$, and half-light radius $r_{\rm e}$
      \citep{sersic1963influence} is created with \texttt{GALSIM},
      with surface brightness profile 
     \begin{equation}
       I(r) = \frac{F}{a_n\,r_{\mathrm{e}}^2}\,
       \exp[-b_n\,(r/r_\mathrm{e})^{1/n}]\,,
    \end{equation}
    where $a_n$ determines the flux normalization and $b_n$ is a known coefficient, which is calculated numerically by \texttt{GALSIM} based on the approximation by \citet{ciotti1999analytical}.
    The generated light profile is convolved with the Moffat profile \citep{moffat1969theoretical}, which is an analytic model for the stellar PSF,
    \begin{equation}
        I(r) = \frac{\beta-1}{\pi r_0^2}[1+(r/r_0)^2]^{-\beta},
    \end{equation}
    where $\beta$ is chosen to be $2.\!\!^{\prime\prime}5$ as a typical value for a seeing-limited PSF and $r_0$ is the full-width at half-maximum (FWHM). 
    We simulate galaxies by randomly selecting entries from the KiDS-1000 weak lensing SOM-gold catalogue and using their fluxes, half-light radii, and ellipticities as input parameters. The ellipticity is incorporated during the profile generation. After convolution with the PSF, each profile is binned according to the KiDS pixel scale of $0.\!\!^{\prime\prime}213/\mathrm{pix}$.
    \begin{figure}
        \centering
        \includegraphics[height=0.27\textheight]{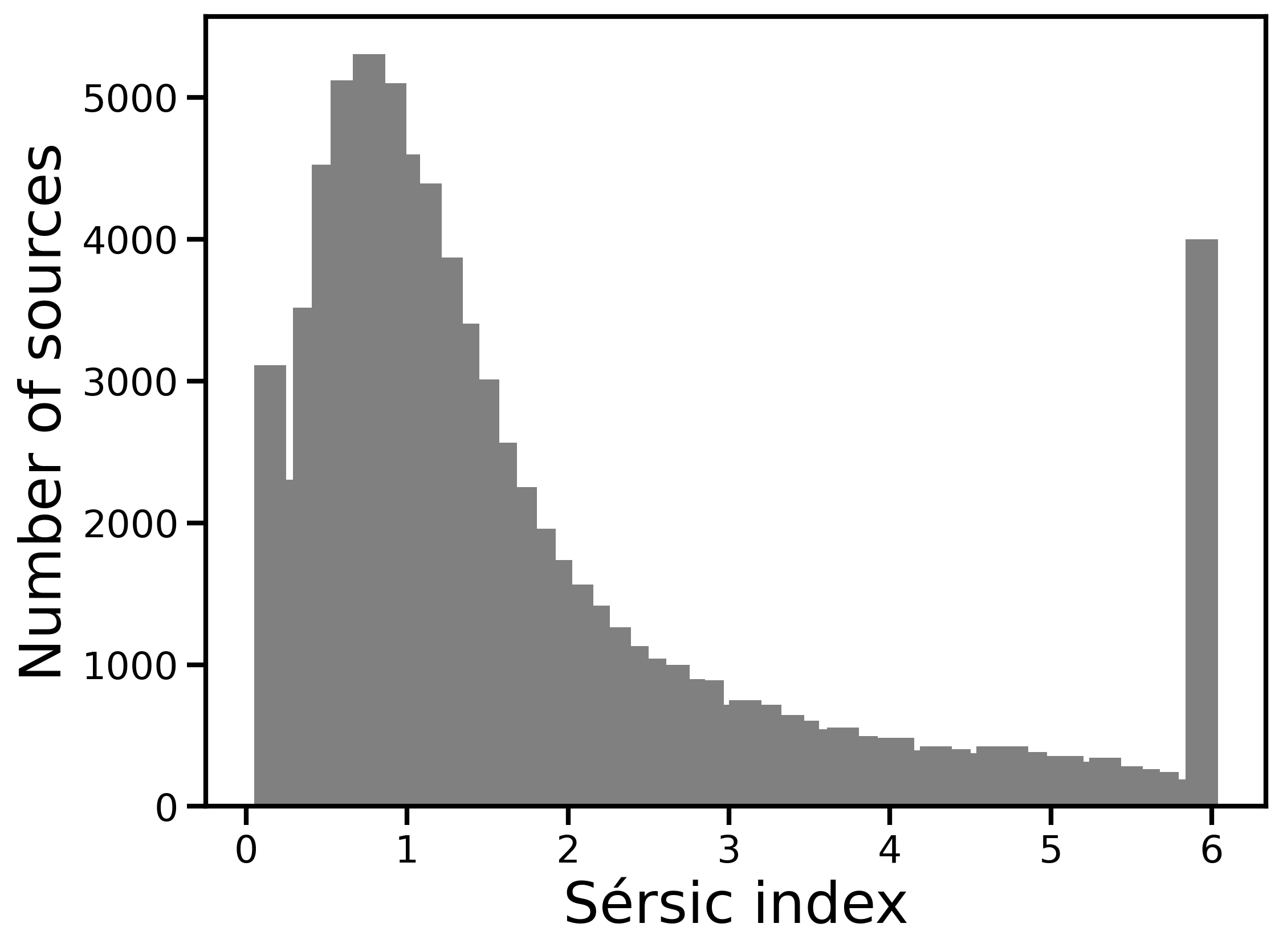}
        \caption[The Sérsic index distribution of the galaxies in COSMOS catalogue.]{Sérsic index distribution of galaxies in the COSMOS catalogue from \citet{hernandez2020constraining}, based on Sérsic profile fits provided by \texttt{GALSIM} \citep{mandelbaum2012, rowe2015galsim}. The sharp peak in the last bin results from a Sérsic index limit in \texttt{GALSIM}; all galaxies with an estimated index greater than 6.2 are grouped into this bin. When sampling Sérsic indices for our simulations, we drew from this distribution while excluding this final bin. For details on the COSMOS catalogue, see \citet{laigle2016cosmos2015}.}
        \label{fig:cosmos_hist}
    \end{figure}
    \item These galaxies are inserted at random but unmasked regions of the corresponding $r$-band detection image, where the two mask images of KiDS-1000, one for the north and the other one for the south, are used to determine the masked regions.
    \item \texttt{SExtractor} (v2.23.1) is run on the source-injected images, which gives 1006 catalogues corresponding to the number of detection images.
    \item These catalogues are compared with the list of injected galaxies in order to assess the retrieval rate of injected galaxies, which gives the detection probability for each image field.
\end{enumerate}
\subsection{Method 1: Cluster regions}
\label{subsec:4.1}
\begin{figure}
    \centering
    \includegraphics[height=0.3\textheight]{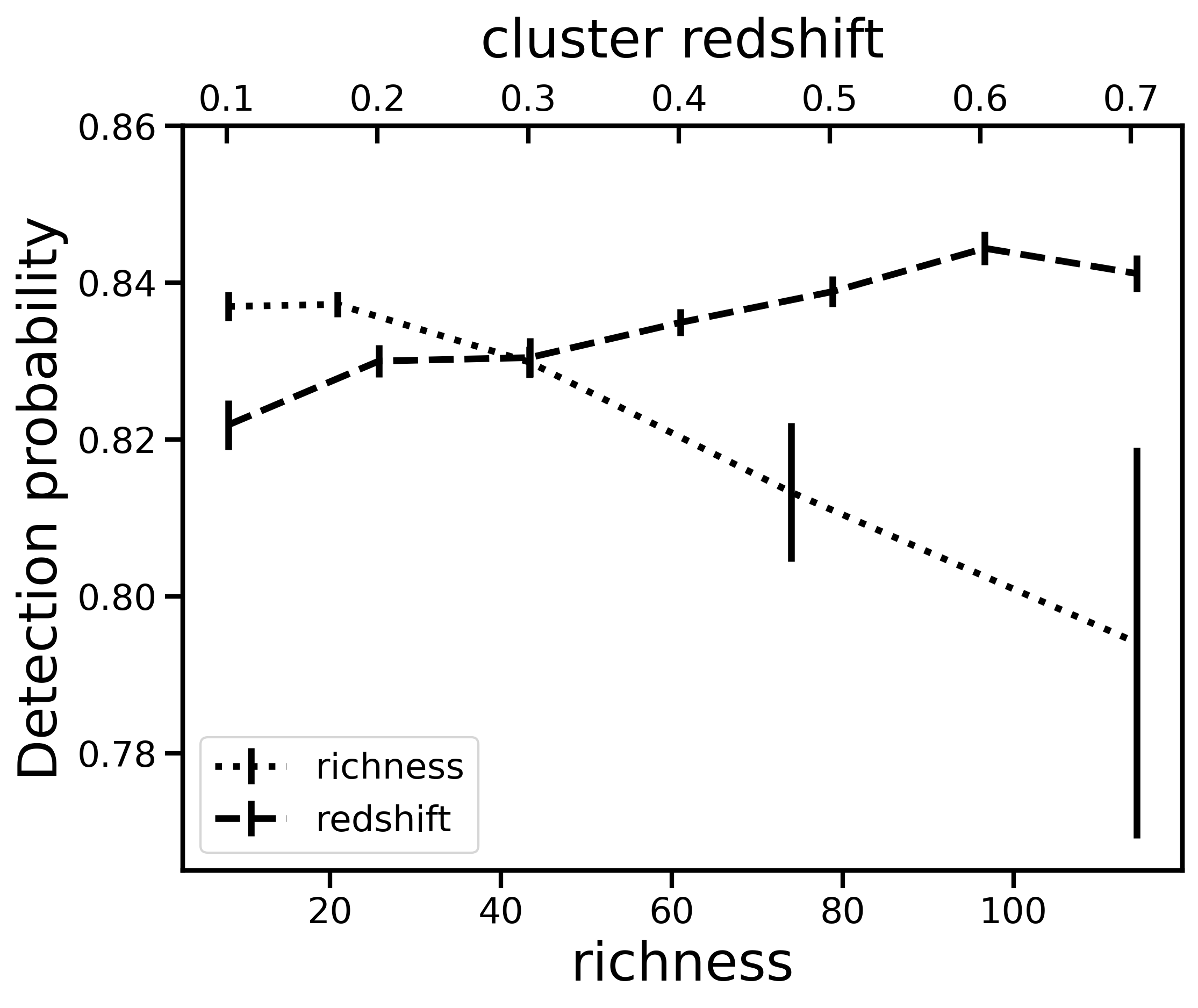}
    \caption{The detection probability as a function of richness
      (bottom axis) and redshift (top axis) of the AMICO-DR3 galaxy clusters, which are shown with dotted and dashed lines, respectively. As there are very limited number of clusters with high richness (i.e., higher than 70) in the catalogue, clusters are divided into 5 logarithmically spaced richness bins and linearly spaced 7 redshift bins. The error bars of the detection probability show the standard errors.}
    \label{fig:cl}
  \end{figure}
Through the complex interplay of dark and baryonic matter across
cosmic time, galaxies formed and aggregated via gravitational
interactions, eventually giving rise to galaxy clusters. These
clusters are now prominent overdense regions in the Universe,
characterised by a high number density of galaxies. As a result,
blending effects are expected to be especially pronounced in cluster
environments, thus in regions where the shear is expected to be
particularly large. The primary goal of this method is to explore how the
detection probability of galaxies varies as a function of cluster
richness and redshift.  

After injecting simulated sources into the images, we assign each
source to its nearest cluster, as identified in the AMICO-DR3
catalogue. A source is excluded from the analysis if its physical
separation from the assigned cluster exceeds the cluster’s virial
radius $R_{200}$.\footnote{We use the $\Lambda$CDM model with
  parameters employed by \citet{maturi2019amico} both in the
  calculation of the virial radii from the mass provided in the AMICO
  catalogue and in determining physical distances from pixels.}

We investigate the detection probability of sources in cluster regions
as a function of cluster richness and redshift. We used the
`intrinsic richness'\footnote{The AMICO-DR3 catalogue provides
  both apparent and intrinsic richness. Since the intrinsic richness
  aligns more closely with conventional definitions of galaxy
  clusters, we adopt it exclusively.} of galaxy clusters in the
AMICO-DR3 catalogue \citep{maturi2019amico}, which is defined as the
sum of the probabilities of galaxies that they belong to the $j$-th
cluster,
\begin{equation}
    \lambda_{j} = \sum_{i=1}^{N_\mathrm{gal}}P_i(j), \hspace{5mm} \begin{cases} m_i<m_*+1.5\\r_i(j)<R_{200}(z_j) \end{cases},
\end{equation}
where $m_*$ is the characteristic magnitude in the Schechter
luminosity function \citep{schechter1976analytic}, which varies with
redshift. We restrict our analysis to clusters with intrinsic richness
$\lambda \geq 5$ to minimise contamination by sparse groups or
spurious detections.

In Fig.~\ref{fig:cl}, we present the detection probability of injected sources within clusters, defined as the recovery rate of injected galaxies in each cluster region. The dotted line shows the average detection probability as a function of cluster richness, and the dashed line shows it as a function of cluster redshift. We find that the detection probability tends to decrease with increasing richness, and increase with redshift. The latter trend may reflect the reduced brightness of contaminating cluster members at higher redshifts, making injected sources easier to detect.

Our method follows a similar approach to that of \citet{kleinebreil_2025}, who injected synthetic galaxies into KiDS tiles using \texttt{GALSIM} to assess detection bias in dense environments. While their analysis used the eRASS1 cluster sample, we used a different cluster catalogue and adopt a simplified framework.  Although both catalogues contain a comparable number of clusters, the eRASS1 sample covers a significantly larger area on the sky ($\sim 1000 \mathrm{deg}^2$ vs. $\sim 450 \mathrm{deg}^2$ in our case), allowing for better statistical robustness and spatial sampling.
They observed that the strength of the object detection bias increases with both cluster redshift and richness, which contrasts with our results, where we find a mild increase of detection probability with cluster redshift. However, a direct comparison is not straightforward due to methodological and data-related differences. In particular, their larger sky coverage allows for a measurement of detection probability as a function of distance from the cluster centre. Moreover, their redshift trend is derived within tomographic and radial bins, while we average over full cluster fields. These differences in scale and technique likely contribute to the divergent trends observed.

Our aim was to model detection probability as a function of both cluster richness and redshift. However, this could not be robustly achieved due to the limited number of clusters in the current catalogue and the strong correlation between these two parameters. Additionally, some clusters span large areas on the sky, such that injected sources may be blended with unrelated field galaxies rather than cluster members. While this issue could be mitigated with colour-based membership information, such data are not available in AMICO-DR3. Despite these limitations, the richness–redshift approach remains a promising direction for future studies. With more comprehensive catalogues and richer cluster data, a more accurate model could be developed. For the present work, we therefore complement this method with a second approach based on the local properties of the inserted sources and their immediate neighbours, which we describe in the next section.

\subsection{Method 2: Source and neighbour properties}
\label{subsec:4.2}
In the second method, we investigate the detection probability of injected sources as a function of their intrinsic properties and those of their nearest neighbouring galaxies. This approach allows us to quantify blending effects in a more general and flexible framework that does not rely on the presence of galaxy clusters.

To capture trends across a range of brightness levels, we analyse each parameter in six linearly spaced magnitude bins of the injected sources: four bins covering the $r$-band magnitude range 20--22, and two bins covering 22--24. Nearest neighbours are restricted to galaxies only. To eliminate stars and unreliable detections from the neighbour sample, we use the \texttt{SG2DPHOT}\footnote{SG2DPHOT: KiDS-CAT star/galaxy classification bitmap based on $r$-band source morphology. Values are: 1 = high-confidence star candidate, 2 = unreliable source, 4 = star based on star/galaxy separation criteria, 0 = other sources (e.g. galaxies). Neighbours with a flag value of 1 or 4 are excluded.} flag provided in the KiDS multi-band catalogue and only retain neighbours with a value of 0.

\begin{figure*}
    \centering
    \includegraphics[height=0.5\textheight]{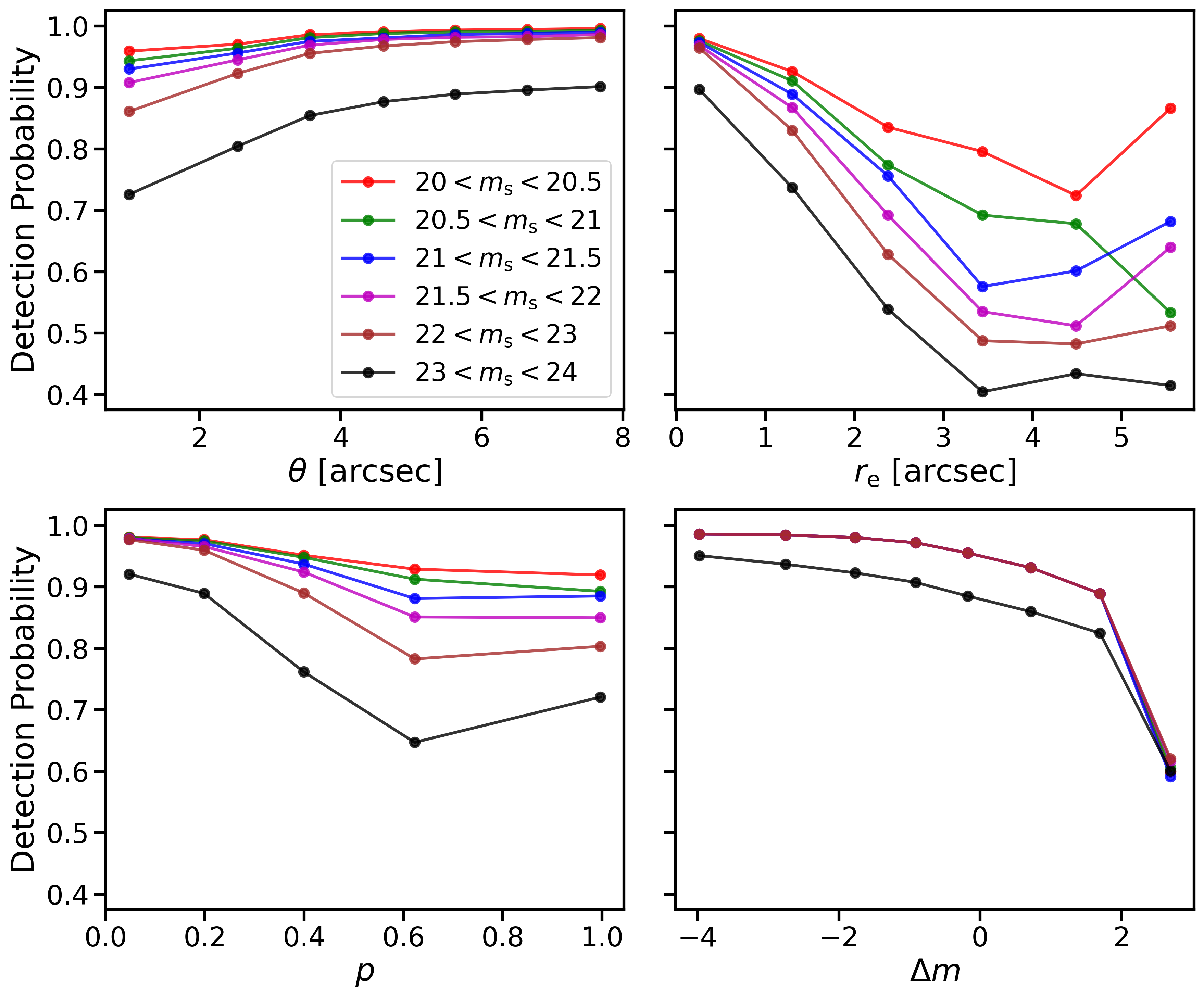}
    \caption{Detection probability of injected source galaxies as a function of four environmental parameters: angular separation to the nearest neighbour (upper left), half-light radius of the nearest neighbour (upper right), proximity parameter (lower left; see Sect.~\ref{subsec:4.2} for definition), and magnitude difference between the injected source and its neighbour (lower right). Curves correspond to different magnitude bins of the injected sources, as indicated in the legend.}
    \label{fig:detprob_m2}
\end{figure*}

Figure~\ref{fig:detprob_m2} shows the detection probability as a function of several source and neighbour parameters. In these plots, subscripts `s' and `n' refer to the properties of the inserted source and its nearest neighbour, respectively.

\begin{itemize}
    \item Separation: The first property investigated is the projected
      separation between the source and its nearest neighbour. As seen
      in the upper left panel of Fig.~\ref{fig:detprob_m2}, the
      detection probability generally increases with separation, as
      expected. However, the dependence is relatively weak, with
      values ranging between approximately 0.75 and 1.0, and the
      behaviour varies across different flux bins. 
    
    \item Half-light radius of the nearest neighbour: The next parameter examined is the half-light radius of the nearest neighbour, shown in Fig.~\ref{fig:detprob_m2} (upper right panel).  In all flux bins, the detection probability decreases with increasing half-light radius, as expected. Compared to the separation, this dependence is more pronounced: detection probabilities can drop below 40\% for sources located near large neighbours. As with the separation, the detailed behaviour differs among flux bins.
    
    \item Proximity parameter: To account for both angular size and separation, we define the `proximity parameter' as
    \begin{equation}
        p = \frac{r_\mathrm{e,s}+r_{\rm e,n}}{\theta}\,,
    \end{equation}
    where $r_\mathrm{e,s}$ and $r_\mathrm{e,n}$ are the half-light radii of the source and the neighbour, respectively, and $\theta$ is their projected separation. This parameter reflects how close two objects appear relative to their combined sizes. Figure~\ref{fig:detprob_m2} (lower left panel) shows the detection probability as a function of this parameter. Similar to the case of separation alone, the dependence is moderate. This suggests that while size and separation both influence detectability, their combined effect through the proximity parameter is still weaker than the influence of the neighbour’s size alone. Moreover, a single functional form cannot capture the behaviour across all flux bins.

  \item Magnitude difference: In Fig.~\ref{fig:detprob_m2} (lower
    right panel), we present the detection probability as a function
    of the magnitude difference between the injected source and its
    nearest neighbour, defined as
    $\Delta m =m\mathrm{_s}-m\mathrm{_{n}}$. Apart from the faintest
    flux bin, the behaviour is consistent across all bins, following a
    smooth and well-defined trend. Due to this robustness and
    simplicity, the magnitude difference is selected as the primary
    parameter for use in subsequent modelling and analysis.
\end{itemize}

\begin{figure}
    \centering
    \includegraphics[width=0.48\textwidth]{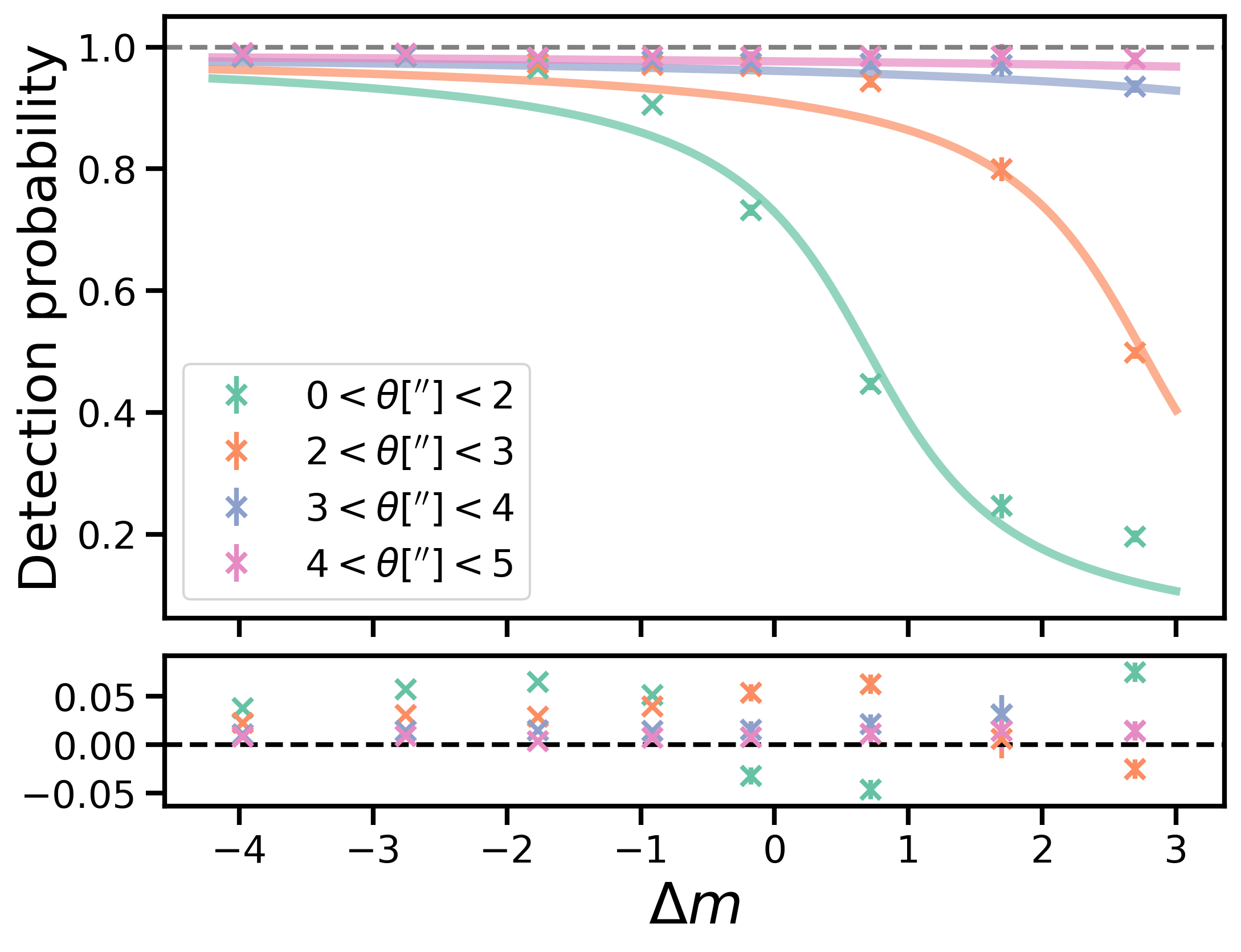}
    \caption{\textit{Upper panel:} The detection probability function
      as a function of the magnitude difference in four separation
      bins between 0 and 5 arcseconds, where the fit function
      (\ref{eq:fitfunc}) for each bin is also plotted. The behaviour
      of the detection probability in higher separation bins coincides
      with the purple line; hence, they are not depicted in this
      figure for clarity. 
      Residuals between the model fit and the data points of each
      separation bin are shown.}
    \label{fig:fitf}
\end{figure}

We examine the detection probability as a function of $\Delta m$ in
four different separation bins, as shown in Fig.~\ref{fig:fitf}. We
adopt a fitting function for the behaviour of the detection
probability as a function of the two driving properties: the magnitude
difference and the separation. To accomplish this, we 
adopt a model function whose choice is
motivated by its ability to provide a good fit to the measurements,
while still being relatively simple,
\begin{equation}
    f(\Delta m, \theta) = 0.5 + \frac{1}{\pi}\, \mathrm{arctan}\left(\frac{b\,\theta - \Delta m}{\delta}\right)\,,
\label{eq:fitfunc}
\end{equation}
where $\Delta m$ is the magnitude difference and $\theta$ is the angular
separation. The parameters $b$ and $\delta$ are
free parameters and are determined via non-linear least-squares
fitting using the \texttt{curve\_fit} function from the \texttt{scipy}
package\footnote{\url{docs.scipy.org/doc/scipy/reference/generated/scipy.optimize.curve\_fit.html}}. The
best-fitting values found are $b = (1.30 \pm 0.03)\,{\rm arcsec}^{-1}$
and
$\delta = 0.50 \pm 0.04$. 

Overall, the model captures the data trends well: residuals are below 5\% for the majority of points and under 10\% for all, except the largest $\Delta m$ bin in the separation range $4<\theta[^{\prime\prime}]<5$.  We used this model function in our subsequent
cosmological analysis, where we aim to quantify how the undetected
sources due to blending can affect cosmic shear measurements.

\subsection{Cosmological analysis}
\label{sec:cosmo}
To quantify the impact of undetected sources due to blending on cosmic shear measurements, we incorporate the constrained detection probability function to assign weights to galaxies. We used these weights as inputs in the computation of the 2PSCFs, enabling a direct comparison of cosmic shear signals with and without accounting for blending effects. Section~\ref{sec:cosmo_1} details the weight assignment procedure, and the shear correlation function measurements are presented in Sect.~\ref{sec:cosmo_2}.

\subsubsection{Assigning weights}
\label{sec:cosmo_1}
For this analysis, we use the \citet{henriques2015galaxy} galaxy catalogue, utilizing the galaxy positions, redshifts, and $r$-band magnitudes. The magnitudes are corrected for both magnification and galactic dust extinction.
We selected the galaxies based on two criteria from the sample:
\begin{itemize}
    \item Redshift: Galaxies are selected in five tomographic redshift bins, matching those used in the KiDS-1000 cosmic shear analysis \citep{asgari_kids}, listed in the second column of Table~\ref{tab:cosmo}.
    \item Flux: Within each redshift bin, galaxies are further
      selected to match the flux distribution observed in
      KiDS-1000.
\end{itemize}
For each galaxy, we compute its detection probability using the model
constructed in Sect.~\ref{subsec:4.2}, based on its projected
separation from the nearest neighbour and the magnitude difference
between the two,
\begin{equation}
    \omega = f(\theta,\Delta m),
\end{equation}
where $\omega$ denotes the detection probability, and $\theta$ and $\Delta m$ are the separation and magnitude difference, respectively. This detection probability is then assigned as a weight to the galaxy and used in the subsequent shear correlation analysis.

To locate the nearest neighbours and to calculate the separations between the galaxies with their neighbours, we used the \texttt{spatial.KDTree}\footnote{\url{docs.scipy.org/doc/scipy/reference/generated/scipy.spatial.KDTree.html}} from \texttt{scipy}, which is a nearest neighbour finding algorithm developed by \citet{maneewongvatana}. 

Although the detection probability weighting alters the effective contribution of galaxies at different redshifts (see Fig.~\ref{figure:weights}), this is a consequence of the underlying detection bias we aim to model. Unlike binary selection schemes, such as those applied by \citet{hartlap2011bias}, our method retains all galaxies and reflects their physical likelihood of being detected. The change in the redshift distribution is thus not imposed by artificial cuts, but emerges naturally from the blending conditions in the data.

\subsubsection{Calculating 2PSCFs}
\label{sec:cosmo_2}

\begin{table}
	\centering
	\caption{Data properties per tomographic redshift bin. The last two columns show the average number of galaxies and the weights assigned to them in each of the 64 field, while the errors show the field-to-field variations.}
	\begin{tabular}{c c c c}
		Bin &$z$-range&  $N_\mathrm{gal}$&$\omega$  \\ [0.5ex]
		\hline
		\hline
		1& $0.1<z\leq0.3$ &$\phantom{2}3707\pm119$& $0.9689\pm0.0004$ \\
		2& $0.3<z\leq0.5$ &$16610\pm277$& $0.9382\pm0.0004$\\
		3& $0.5<z\leq0.7$&$40133\pm434$& $0.9196\pm0.0003$\\
		4& $0.7<z\leq0.9$&$69378\pm574$& $0.9185\pm0.0003$\\
		5&$0.9<z\leq1.2$&$94764\pm680$& $0.9136\pm0.0003$\\
		\hline
	\end{tabular}
	\label{tab:cosmo}
\end{table}
The ray-tracing results from the Millennium Simulation provide the components of the Jacobian matrix on 64 redshift slices, each represented by a $4096\times4096$ pixel grid. From these data, we compute key lensing quantities: convergence, which traces the matter distribution, and the reduced shear, which serves as the input for computing the shear correlation functions.

Since the galaxy positions in the \citet{henriques2015galaxy}
catalogue are derived from dark matter halo coordinates, they are
given in angular units (radians), not in pixel coordinates. To
associate lensing quantities with galaxies, we project the galaxy
positions onto the pixel grid and assign them the reduced
  shear values of the nearest grid points.

A summary of the data used in the shear correlation function calculations is provided in Table~\ref{tab:cosmo}. The average number of galaxies per field and their corresponding mean weights are listed in the third and fourth columns. We observe that the average weight tends to decrease with redshift, reflecting the increasing difficulty of detecting high-redshift galaxies that are fainter and more likely to be blended with brighter neighbours.

\begin{figure}
    \centering
    \includegraphics[width=0.47\textwidth]{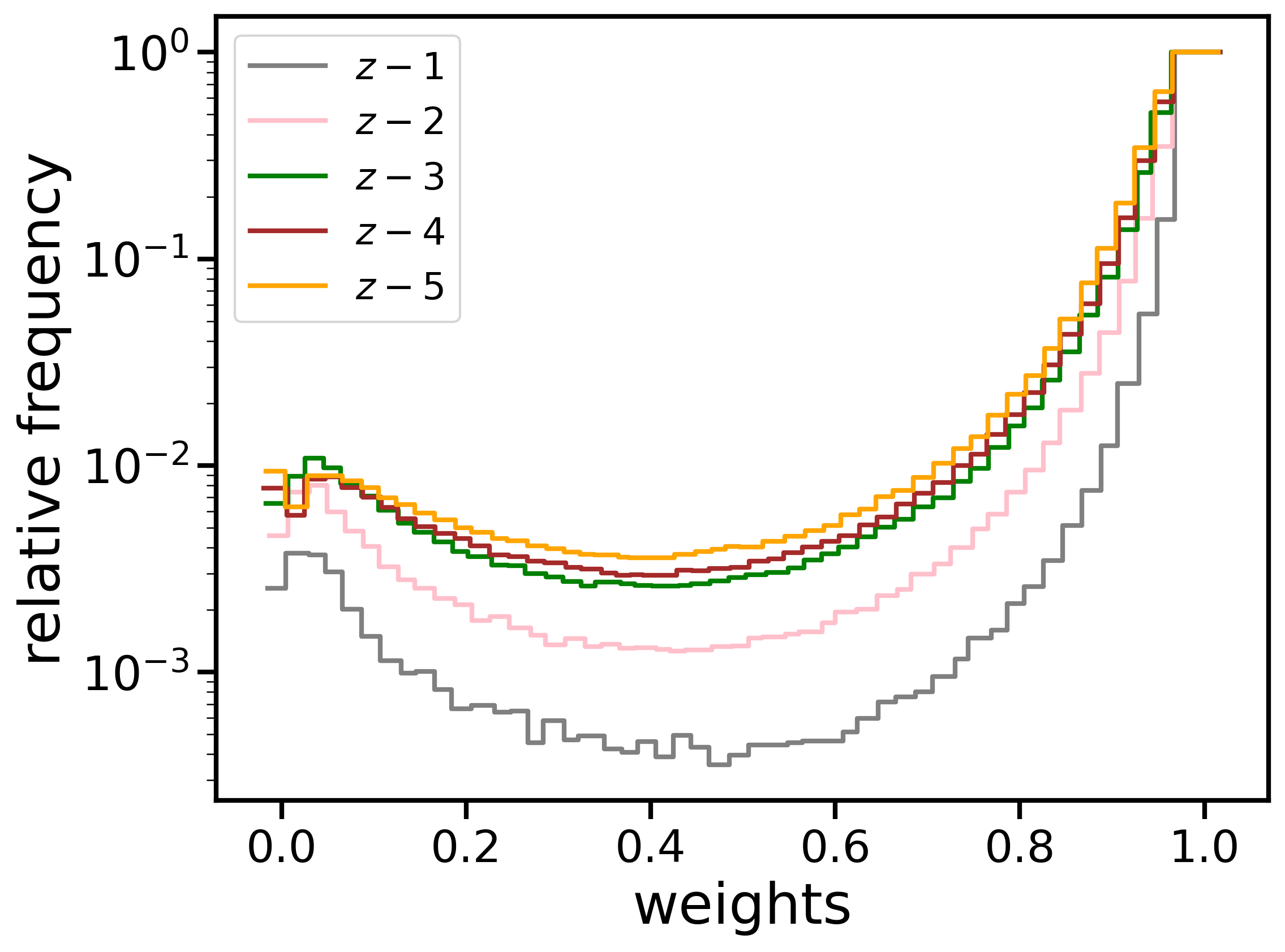}
    \caption{Distribution of the calculated detection probabilities (i.e. assigned weights) of galaxies in each $z$-bin. A primary peak is visible near unity, while a secondary peak appears near zero, particularly in the lower redshift bins.}
    \label{figure:weights}
  \end{figure}
Figure~\ref{figure:weights} shows the distribution of detection
probabilities (assigned weights) across redshift bins. While the
majority of galaxies have weights close to unity, a secondary peak
near zero is also visible, especially in lower redshift bins. This
feature is attributed to the larger isophotal sizes of low-redshift
galaxies, which increase the likelihood of blending with nearby
sources. 

\begin{figure*}
    \centering
    \includegraphics[width=0.9\textwidth]{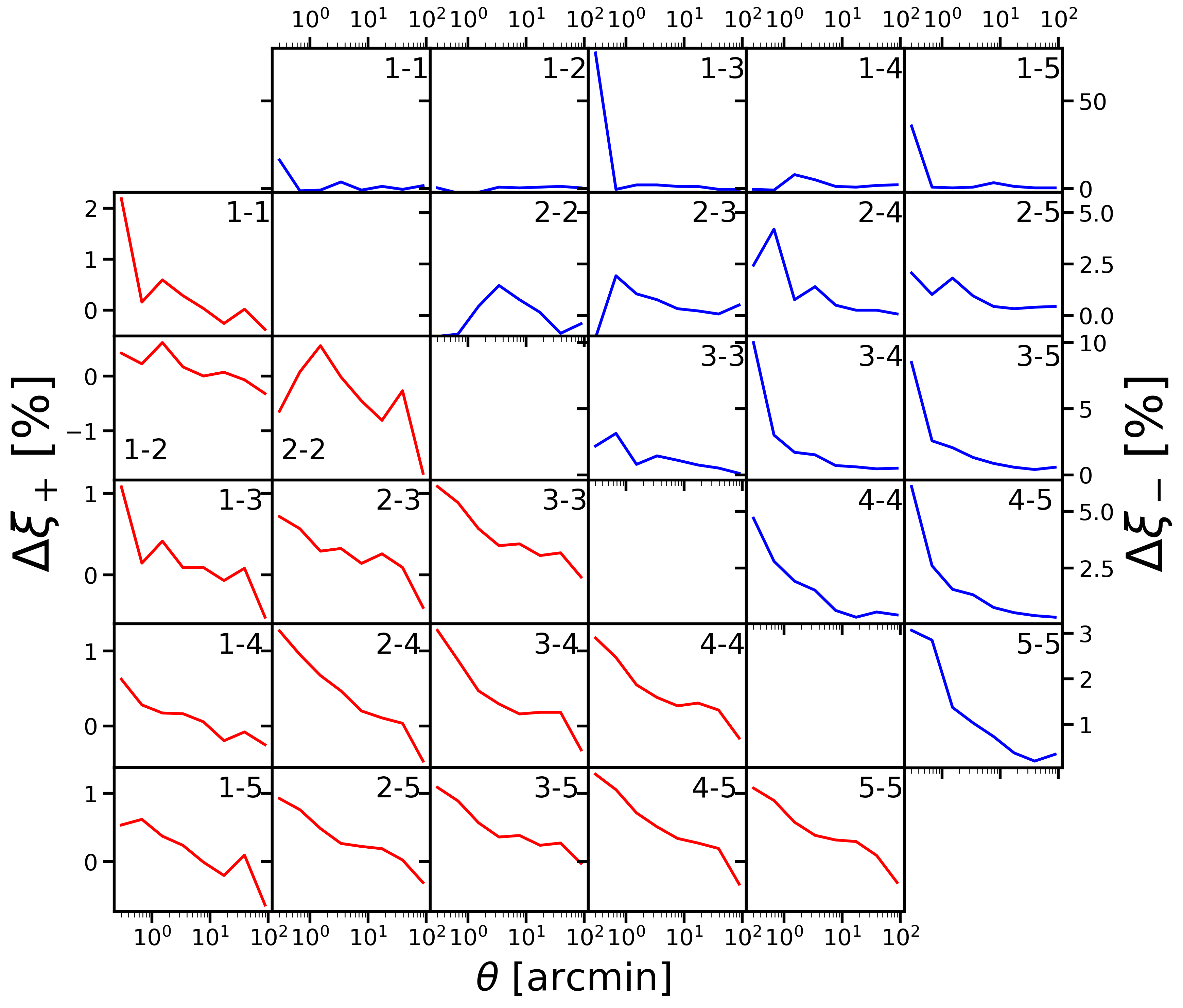}
    \caption{Bias (Eq.~\ref{eq:bias}) in the 2PSCFs. The red colour represents the bias in $\xi_+$, while the blue colour shows the bias in $\xi_-$. Each plot is labelled with the corresponding tomographic redshift bin pair. 
    }
    \label{fig:farkallcorrs}
\end{figure*}

We perform the calculation of the 2PSCFs with \texttt{TreeCorr}
\citep{treecorr} in eight logarithmically spaced angular separation
bins between $0.\!\!^{\prime}2$ and $120^{\prime}$, matching the
binning strategy of the 
KiDS-1000 cosmic shear analysis. This choice optimizes the balance
between signal-to-noise ratio and statistical robustness: using too
few bins can obscure scale-dependent effects, while too many would
introduce noise and reduce statistical significance. Initially, the
correlation functions are computed without applying any weights. The
analysis is then repeated with the galaxy detection probabilities
incorporated as multiplicative weights.

In Fig.~\ref{fig:farkallcorrs}, we present the bias in the 2PSCFs due to blending, defined analogously to \citet{hartlap2011bias} as
\begin{equation}
    \Delta\xi_{\pm} = \frac{\xi_{\pm,\mathrm{nw}} - \xi_{\pm,\mathrm{w}}}{\xi_{\pm,\mathrm{nw}}}\,,
\label{eq:bias}
\end{equation}
where the subscript of `nw' denotes the non-weighted and `w' the
weighted correlation functions. The relative bias is considerably
smaller in $\xi_+$ compared to the bias in $\xi_-$, and decreases with
increasing angular separation, indicating that the blending-induced
bias predominantly affects small physical scales. This is consistent
with theoretical expectations: the Bessel function $J_4$ in
Eq.~\eqref{eq:corrbessels}, which defines the filtering of the
power spectrum for $\xi_-$, peaks at $\ell\theta \approx 5$, making
$\xi_-$ more sensitive to smaller scales. Conversely, $\xi_+$ is
dominated by large-scale modes due to the peak of $J_0$ at $\ell\theta
= 0$. The particularly large bias in $\xi_-$ for the lowest
redshift bin stems from the extremely small amplitude of the
correlation function itself in this regime.

\subsubsection{Cosmological parameters}
To assess the impact of this detection bias on cosmological inference,
we perform an analysis based on the likelihood 
\begin{equation}
    \mathfrak{L} \propto \mathrm{exp}\left(
      -0.5[\boldsymbol{\xi}-\boldsymbol{\xi}_\mathrm{m}]^{\rm t}\mathrm{C}^{-1}[\boldsymbol{\xi}-\boldsymbol{\xi}_\mathrm{m}]\right)\,,
\end{equation}
where the data vector of observed correlation functions is given by
$\boldsymbol{\xi}=[\xi_+(\theta_1), ..., \xi_+(\theta_8),
\xi_-(\theta_1), ..., \xi_-(\theta_8)]^{\mathrm{t}}$. 
These functions are measured using the \citet{henriques2015galaxy}
catalogue in the same eight angular separation bins between
$0.\!\!^{\prime}2$ and $120^{\prime}$. The model prediction vector
$\boldsymbol{\xi}_\mathrm{m}$ is computed from a theoretical shear
power spectrum generated with \texttt{pyccl}
(v3.0.1)\footnote{\url{https://ccl.readthedocs.io}}. The covariance
matrix $\mathrm{C}$ is adopted from the KiDS-1000 cosmic shear
analysis \citep{asgari_kids}. 

Table \ref{tab:priors} lists the cosmological parameters and priors
used in the likelihood analysis. The baryonic matter density parameter
$\Omega_\mathrm{b}$ is fixed, while the cold dark matter density
$\Omega_\mathrm{c}$ and the amplitude of the power spectrum $\sigma_8$
are allowed to vary. The likelihood is first evaluated using
unweighted correlation functions, followed by one that
includes the detection probability weights.

We find a fractional bias of $\sim2.35\%$ in $\Omega_\mathrm{c}$ and $\sim0.66\%$ in $\sigma_8$, resulting in an overall bias of $\sim1.88\%$ in the derived parameter $S_8 = \sigma_8\sqrt{\Omega_\mathrm{m}/0.3}$. It is important to note that this estimate does not include parameter degeneracies, nor does it account for additional systematic effects such as baryonic feedback or intrinsic alignments.
\begin{table}
    \centering
    \caption{Cosmological parameters and their priors used in fiducial cosmology. The cold dark matter and the baryonic matter density parameter are represented by $\Omega_\mathrm{c}$ and $\Omega_\mathrm{b}$, respectively, while $\sigma_8$ denotes the normalization of the power spectrum, $n_\mathrm{s}$ is the spectral index of the primordial power spectrum, and $h$ is the dimensionless Hubble parameter.}
    \begin{tabular}{c c}
        parameter & prior \\ 
        \hline \hline
        $\Omega_\mathrm{c}$ & [0.1, 0.5] \\
        $\sigma_8$ & [0.5, 1.1] \\
        $\Omega_\mathrm{b}$ & 0.045  \\
        $n_\mathrm{s}$ &  1.0\\
        $h$ & 0.73 \\
        \hline
    \end{tabular}
    \label{tab:priors}
\end{table}

\section{Conclusion}
\label{sec:conclusion}
In this study, we investigated the impact of galaxy non-detections on
cosmic shear measurements, first investigated by \citet{hartlap2011bias} but 
since then received little attention. This effect adds to that of blending
sources, which renders shear measurements less reliable and which has
been studied in detail before \citep{hoekstra2017study,
  mandelbaum2018weak, samuroff2018dark, martinet_2019}. With the increasing
demand for precise systematic control in Stage-IV cosmological
surveys, we revisit and expand upon the considerations of
\citet{hartlap2011bias}, deepening our understanding of galaxy
detection probabilities under varying conditions. In particular,
instead of discarding galaxy pairs based on a plausible but ad-hoc
separation criterion, we empirically determine the detection
probability based on KiDS-1000 data. 

We initially explored detection biases in the vicinity of galaxy
clusters. The detection probability shows a mild increase with cluster
redshift and a possible decrease with cluster richness, although the
trends are not strongly significant and are accompanied by large
uncertainties at higher richness, which can likely be attributed to
the limited number of clusters in the catalogue and the strong
correlation between cluster parameters. 

To overcome this, we pursued an alternative strategy by analyzing the
detection probability of inserted sources as a function of their
intrinsic properties and those of their neighbouring galaxies, across
different flux bins. We found that the detection probability exhibits
a consistent dependence on the $r$-band magnitude difference between a
source and its nearest neighbour, regardless of the source
flux. Interestingly, this finding contrasts with the results of
\citet{samuroff2018dark}, who observed a weaker dependence on
magnitude difference relative to angular separation. This discrepancy
highlights the importance of first determining whether galaxies are
detectable at all before assessing the precise impact of blending.

We constrained our detection probability function depending
both on the magnitude difference and the separation. In crowded galaxy fields, blending can occur not only from the closest neighbouring galaxy but also from multiple nearby sources simultaneously. Additionally, the second-nearest neighbour, the galaxy closest after the nearest one, can contribute to blending effects. Our method captures these complexities as well, as we average over all possible pairs of galaxies. 

To quantify this bias in cosmic shear surveys, we employed our
detection probability function to assign weights to galaxies, which we
utilized in computing shear correlation functions. A key advantage of
this weighting scheme is that it preserves the full galaxy sample
while down-weighting less reliably detected sources. As shown in
Fig.~\ref{fig:farkallcorrs}, the bias is most prominent on small
angular scales and in the $\xi_-$ correlation function. Our likelihood
analysis demonstrated that this bias can lead to a fractional shift of
$1.88\%$ in the $S_8$ parameter, underscoring its relevance for
precision weak lensing studies. 

It is important to realise that this source non-detection effect
provides a lower bound on the impact of blending on cosmic shear
measurements. Detected, but blended galaxies typically are assigned
lower weights due to the increased uncertainty of shear estimates.

In summary, we find that the flux ratio between a galaxy and its
neighbours, along with their angular separation, are the key
determinants of detectability. The detection probability function
derived in this work provides a valuable tool for future efforts to
mitigate detection-related systematics -- either through the development
of new algorithms or by enhancing existing ones, such as
\texttt{Metadetection} \citep{sheldon2020}, to more effectively
identify and correct for blended sources.

\begin{acknowledgements}
  We would like to thank Sven Heydenreich and Laila Linke for useful
  discussions during the project. EG acknowledges the support from the Deutsche Forschungsgemeinschaft (DFG) SFB1491. TS acknowledges support from the Austrian Research Promotion Agency (FFG) and the Federal Ministry of the Republic of Austria for Climate Action, Environment, Mobility, Innovation and Technology (BMK) via grants 899537, 900565, and 911971. The data used in this work are based
  on observations made with ESO Telescopes at the La Silla Paranal
  Observatory under programme IDs 177.A-3016, 177.A-3017, 177.A-3018
  and 179.A-2004, and on data products produced by the KiDS
  Consortium.  The KiDS production team acknowledges support from the
  DFG, ERC, NOVA and NWO-M grants; Target;
  the University of Padova, and the University Federico II
  (Naples). This work was supported by a grant of the German Centre of
  Cosmological Lensing, hosted at Bochum University.
 The Millennium Simulation data sets were
  constructed as part of activities of the German Astrophysical
  Virtual Observatory.

\end{acknowledgements}

%
%

\bibliography{References} 

%

\label{LastPage}
\end{document}